\documentclass[11pt,a4paper]{article}
\usepackage{etex,amsmath,amsfonts,amssymb,amsthm,tikz,a4wide,mathrsfs}
\pdfoutput=1
\numberwithin{equation}{section}
\usetikzlibrary{decorations.pathmorphing}

\begin{document}

\begin{center}

{\Large Conformal isometry of the lukewarm}

{\Large Reissner-Nordstr\"om-de Sitter spacetime}

\vspace{8mm}

{\large Helgi Freyr R\'unarsson}\footnote{helgi.runarsson@gmail.com}

\vspace{4mm}

{\sl Fysikum, Stockholms Universitet,} 

{\sl S-106 91 Stockholm, Sweden}

\vspace{6mm}

{\bf Abstract:}

\end{center}

\noindent It is known that the extremal Reissner-Nordstr\"om black hole possesses a discrete conformal isometry that exchanges the black hole horizon with infinity.
It is also known that the Reissner-Nordstr\"om-de Sitter spacetime posseses a similar discrete conformal isometry which exchanges the event horizon of the black hole with the cosmological horizon.
In this short paper we will continue this line of thought and extend the conformal isometry of the latter spacetime and give an unphysical interpretation of the negative root.

\section{Introduction}

\noindent A spacetime, described by the metric $g_{ab}$, is said to possess a discrete conformal isometry if there exists a map of the spacetime itself under which distances can change but only by an overall factor, such that angles are preserved.

As Couch and Torrence showed\cite{RNconfiso}, the extremal Reissner-Nordstr\"om black hole possesses a discrete conformal isometry that exchanges the black hole horizon and infinity.
Later Br\"annlund showed\cite{RNdSconfiso} that the lukewarm Reissner-Nordstr\"om-de Sitter spacetime possesses a similar discrete conformal isometry that exchanges the outer black hole horizon with the cosmological horizon.

We will start by introducing the Reissner-Nordstr\"om-de Sitter spacetime and define what a lukewarm spacetime means.
The Reissner-Nordstr\"om-de Sitter spacetime is described by the metric

\begin{align}
    ds^2 &= -V(r)dt^2 + V^{-1}(r)dr^2 + r^2d\Omega_2^2
    \label{eqn:RNdSmetric}
\end{align}
where
\begin{align}
    \begin{alignedat}{2}
    V(r) &= \frac{1}{r^2}\left( r^2-2mr+e^2-\frac{\lambda}{3}r^4 \right) \\
    &= \frac{1}{r^2}\left( r-r_a \right)\left( r-r_b \right)\left( r-r_c \right)\left( r-r_d \right),
    \label{eqn:Vofr}
    \end{alignedat}
\end{align}
where $m$ is the mass of the black hole, $e$ the charge, $\lambda$ the cosmological constant, $r_a$ the inner black hole horizon, $r_b$ the outer black hole horizon, $r_c$ the cosmological horizon and $r_d$ is the negative root and the following holds
\begin{align}
    r_d<0<r_a<r_b<r_c.
    \label{eqn:orderofroots}
\end{align}

If we restrict ourselves to the case of a lukewarm spacetime, when the Hawking temperatures of the outer black hole horizon and the cosmological horizon are equal, $V(r)$ takes on the simple form\cite{lukewarmbh}

\begin{align}
    V(r) &= \frac{r^4-\left( r-m \right)^2}{r^2}.
    \label{eqn:Vofrlukewarm}
\end{align}
It is now easy to solve explicitly for the roots.

\section{Conformal isometry}

\noindent In order to see the conformal isometry of the lukewarm spacetime more clearly, we define a new coordinate $x$ by

\begin{align}
    x(r) &= \frac{r-r_b}{r_c-r}\frac{r_c}{r_b}. 
    \label{eqn:xofr}
\end{align}
When the spacetime is given in terms of this new coordinate, the inversion
\begin{align}
    x &\rightarrow x'=1/x
    \label{eqn:inversion}
\end{align}
is a discrete conformal isometry of the spacetime.
That is, under the inversion the following equation holds

\begin{align}
    ds^2 &= \Omega(x)d\overline{s}^2,
\end{align}
where $ds^2$ is the metric in terms of $x'$, $d\overline{s}^2$ is the metric in terms of $x$ and $\Omega(x)$ is given by
\begin{align}
    \Omega(x) &= \frac{xr_b+r_c}{xr_c+r_b}. 
\end{align}
Now we can discuss what was mentioned in the abstract and introduction, which hypersurfaces are exchanged under this conformal isometry.
From the form of $x(r)$ we see that the following equations hold
\begin{align}
    x(r_b)x(r_c) = 1, && x(m)x(\infty) = 1, && x(r_a)x(r_d) = 1, 
    \label{eqn:exchanges}\\
    x(2m) = 1, && x(0) = -1.
\end{align}
That is:
$r_b$ and $r_c$ are mapped into each other,
$r=m$ and $r=\infty$ are mapped into each other,
$r_a$ and $r_d$ are mapped into each other,
$r=2m$ and $r=0$ are mapped into themselves
under the conformal isometry.

At $r=m$ the four dimensional Weyl tensor vanishes which is what happens at $\mathscr{I}$ ($r=\infty$).
This is easily seen from the form of the squared Weyl tensor,

\begin{align}
    C_{abcd}C^{abcd} &= \frac{48(r-m)^2m^2}{r^8}.
    \label{eqn:WeylSq}
\end{align}
Since this is a conformally invariant statement, any hypersurface that is exchanged with $\mathscr{I}$ must have this property.
The hypersurface $r=2m$ is a photon sphere\cite{photonsphere}.

It is worth mentioning that the choice of $x(r)$ in equation \eqref{eqn:xofr} was found by insisting that

\begin{align}
    x(2m) = 1, && x(r_b) = 0, && x(r_c) = \infty.
    \label{eqn:xofrdefinition}
\end{align}
In fact, any definition that insists that one of the three pairs, ($r_b$, $r_c$), ($r=m$, $r=\infty$) or ($r_a$,$r_d$), are mapped into zero and infinity and $r=2m$ is mapped into $1$ or $r=0$ is mapped into $-1$ (that is, they are the fixed points of the mapping) will give the same results as we have shown here.
Therefore the choice of $x(r)$ above is in no way special.

We can visualize these results in a Carter-Penrose diagram for the conformally extended lukewarm spacetime.
We follow Walker's algorithm\cite{walker} and glue together the 4 different blocks of the spacetime.
These four blocks are bounded by
$r=\infty$ and $r_c$,
$r_c$ and $r_b$,
$r_b$ and $r_a$,
$r_a$ and $r=0$.

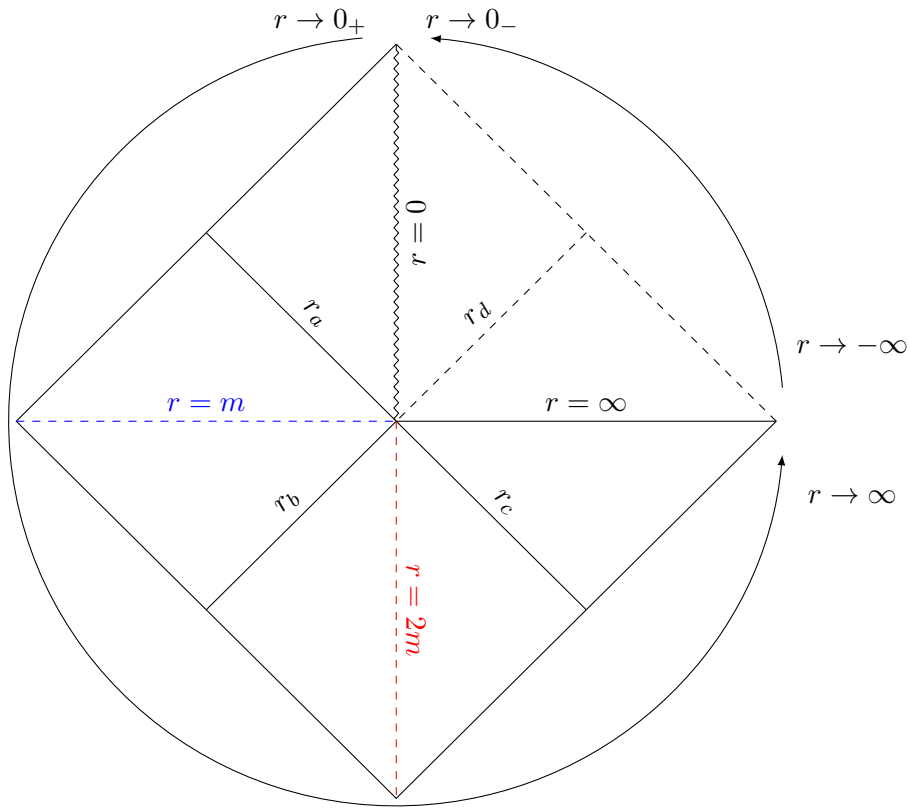
\begin{figure}
    \begin{center}
        \begin{tikzpicture}
    \draw [line join=round,decorate, decoration={zigzag,segment length=4,amplitude=.9,post=lineto,post length=2pt}]  (5,0) -- (5,5) node[sloped,midway,below]{$r=0$};
    \draw (5,0) -- (2.5,2.5) node[sloped, midway, above]{$r_a$};
    \draw (2.5,2.5) -- (5,5);
    \draw (5,0) -- (2.5,-2.5) node[sloped,midway,above]{$r_b$};
    \draw (2.5,-2.5) -- (0,0) -- (2.5,2.5);
    \draw (2.5,-2.5) -- (5,-5) -- (7.5,-2.5);
    \draw (7.5,-2.5) -- (5,0) node[sloped,midway,above]{$r_c$};
    \draw (5,0) -- (10,0) node[sloped,midway,above]{$r=\infty$};
    \draw (7.5,-2.5) -- (10,0);
    \draw[dashed] (10,0) -- (5,5);
    \draw[dashed] (7.5,2.5) -- (5,0) node[sloped,midway,above]{$r_d$};
    \draw[dashed,color=red] (5,0) -- (5,-5) node[sloped,midway,above]{$r=2m$};
    \draw[dashed,color=blue] (0,0) -- (5,0) node[sloped,midway,above]{$r=m$};
    \draw[-latex] (4.5555,5.081) arc (95:355:5.1);
    \draw (4,5.3) node{$r\rightarrow0_+$};
    \draw (11,-1) node{$r\rightarrow\infty$};
    \draw[-latex] (10.081,0.4445) arc (5:85:5.1);
    \draw (6,5.3) node{$r\rightarrow0_-$};
    \draw (11,1) node{$r\rightarrow-\infty$};
\end{tikzpicture}
    \end{center}
    \caption{Carter-Penrose diagram for the conformally extended Reissner-Nordstr\"om-de Sitter spacetime. The circles around the diagram indicate in which direction $r$ grows. The dashed lines play a role in the conformal isometry of the spacetime as described above.}
    \label{fig:RNdSpenrose}
\end{figure}

However, we can take inspiration from statistical mechanics and negative temperature systems and extend the diagram beyond infinity.
That is, we extend the first and fourth blocks and glue them together along $r_d$ and let $r$ go between minus infinity and zero in this new region.
The result is shown in figure \ref{fig:RNdSpenrose}.
 
We see that the inversion acts as a mirroring of the regions on either side of the fixed points of the inversion, $r=2m$ and $r=0$.

Therefore we see that the negative root, $r_d$, has an unphysical interpretation since it is conformal to the inner horizon of the black hole, $r_a$, as we saw in equation \eqref{eqn:exchanges}.

\section{Conclusions and acknowledgements}

The lukewarm Reissner-Nordstr\"om-de Sitter spacetime possesses a discrete conformal isometry that can be extended beyond infinity to negative values of $r$.
This conformal isometry exchanges the outer black hole horizon and the cosmological horizon, the inner black hole horizon and the negative root as well as the two hypersurfaces at $r=m$ and $r=\infty$.

I would like to thank my supervisor Ingemar Bengtsson for all the help, since without him, this work would not have been completed.

\bibliographystyle{h-physrev5.bst}
\bibliography{references}

\begin{thebibliography}{1}

\bibitem{RNconfiso}
W.~Couch and R.~Torrence,
\newblock Gen. Rel. Grav. {\bf 16}, 789 (1984).

\bibitem{RNdSconfiso}
J.~Br\"annlund,
\newblock Gen. Rel. Grav. {\bf 36}, 883 (2004).

\bibitem{lukewarmbh}
L.~J. Romans,
\newblock Nucl. Phys. {\bf B383}, 395 (1992).

\bibitem{photonsphere}
C.-M. Claudel, K.~S. Virbhadra, and G.~F.~R. Ellis,
\newblock J. Math. Phys. {\bf 42}, 818 (2001).

\bibitem{walker}
M.~Walker,
\newblock J. Math. Phys. {\bf 11}, 2280 (1970).

\end{thebibliography}

\end{document}